\documentstyle[12pt]{article}
\def\a{\alpha}
\def\b{\beta}
\def\g{\gamma}
\def\d{\delta}
\def\l{\lambda}
\def\m{\mu}
\def\R{{\cal R}}
\def\K{{\cal K}}
\newcommand{\D}{\mbox{$\Delta_+\,$}}
\newcommand{\bn}{\begin{equation}}
\newcommand{\ed}{\end{equation}}

\newcommand{\Rmt}{ universal $R$--matrix }

\newtheorem{definition}{Definition}[section]

\newtheorem{theorem}{Theorem}[section]
\newtheorem{corollary}{Corollary}[section]
\newcommand{\hsp}{\mbox{$\hspace{.5in}$}}

\begin{document}
\title{Gauss decomposition of trigonometric R-matrices}
\author{S.M.Khoroshkin \\
ITEP, Bol'shaya Cheremushkinskayas 25,\\
117259 Moscow, Russia \\
A.A.Stolin \\
Department of Mathematics, Royal Institute of Technology,\\
S--10044 Stockholm, Sweden \\
V.N.Tolstoy \\
Institute of Nuclear Physics, Moscow State University,\\
119899 Moscow \\}
\date{}
\maketitle
\begin{abstract}
The general formula for the universal R-matrix for quantized nontwisted
 affine algebras by Khoroshkin and Tolstoy is applied for zero central charge
 highest weight modules of quantized affine algebras. It is shown how the
 universal $R$-matrix produces the Gauss decomposition of trigonomitric
 $R$-matrix in tensor product of these modules. Explicit calculations for
 the simplest case of $A_1^{(1)}$ are presented. As a
 consequence new formulas for the trigonometric
 $R$-matrix with a parameter in tensor product of
 $U_q(sl_2)$-Verma modules are obtained.
\end{abstract}
 \setcounter{section}{-1}
\section{Introduction}
The most known technique for calculating trigonometric solutions of the
 Yang--Baxter equation (with a parameter) looks as follows.

One can fix a quantum deformation $U_q(g)$ of the universal enveloping algebra
 of a simple Lie algebra $g$ and write down the trigonometric solutions of the
 Yang--Baxter equation in fundamental representations of $U_q(g)$ which are all
 known (see \cite{J1}). Then the $R$-matrices in other
 finite-dimensional representations of $U_q(g)$ can be
 obtained by so called fusion procedure \cite{KS},
 \cite{KRS}, \cite{J2}
which is based on the analysis  of
 $q$-analogs of the Young symmetrizes in tensor
 productes of fundamental representations \cite{Ch1},
 \cite{Ch2}.

In  alternative approach one can apply the universal $R$-matrix \cite{Dr},
\cite{Dr1} to tensor product of concrete representations. For the purpose
 of parametric trigonometric solutions of the Yang--Baxter equation
 one should use the \Rmt for quantized affine algebras. An explicit formula
 for \Rmt for quantized nontwisted affine algebras was found in \cite{KT2}.
 It has multiplicative form and can be presented in terms of infinite
 product over the set of all positive root of the corresponding affine
 algebra with the factors being some $(q)$-exponential functions of
$U_q(\widehat{g}\otimes \widehat{g})$. The complicated form of these
infinite products restricts the number of applications of the universal
$R$-matrix. However, some low dimensional examples are given in \cite{KT2},
 \cite{KT3}, \cite{yzz}.

The aim of this paper is to remove some psychological obstacles and to
demonstrate the technique of working with the universal $R$-matrix in tensor
 product of some zero charge highest weight representations of quantized
affine  algebras.

First we note that due to the general structure of normal (or convex in
 terminology, suggested by V.G.Kac) ordering of positive roots of affine
 Lie algebra the \Rmt  can be naturally divided into three parts which
 produce  generalized Gauss decomposition of
 $R$-matrix in concrete representations.

Next we demonstrate in $A_1^{(1)}$ case how to calculate these three parts
in tensor products of the representations, induced from Verma or finite
 dimensional modules of finite $U_q(sl_2)$ (more exactly, for corresponding
 evaluation representations of ${U'}_q(\widehat{sl}_2)$).
 Moreover, we reexpress the entries of the Gauss decomposition in terms of
$U_q(sl_2)\otimes U_q(sl_2)$ generators with coefficients being degenerated
 basic hypergeometric functions of Cartan elements.
 As a consequence we obtain the general form of Gauss decomposition of
 trigonometric $R$-matrix in tensor product of $sl_2$-Verma modules
 (and of finite dimensional modules as factors of Verma modules).
These examples show that a field of applications of the universal $R$-
matrix is not covered by fusion technique.

It is not useless to look at the structure of presented Gauss factors of
 trigonometric $R$-matrix. The nilpotent parts have very simple
$q$-exponential structure with coeffitients being rational function of
spectral parameter usual for trigonometric $R$-matrix. The diagonal
 component can be divided into three different parts. The first gives the
 scalar factor \cite{Dr}, \cite{KT2} being the solution of one-dimensional
 $q$-difference Knizhnik-Zamolodchikov equation \cite{FR}.
The two others have
 rational coefficients for finite dimensional
 representations but one of them is not rational for
 Verma modules with noninteger highest weights.  In
 this case the $R$-matrix cannot be expressed in terms
 of rational functions of spectral parameter.
\vskip 0.3cm

{\bf Acknowlegements.}
The authors are thankful to the Academy of Science of Sweden for the
hospitality of Royal Institute of Technologies where the first steps of the
 work were done. The first and the third authors are also especially thankful
 to Professor C.M.Ringel for the kindly atmosphere in SFB 343 of
 Mathematical Departement of Bielefeld University where the work was
 completed.

\section{Notations}

Let $\hat{g}$ be nontwisted affine Lie algebra  with symmetrizable
Cartan matrix $A=(a_{ij})$
($A^{sym}=(a_{ij}^{sym})$ is the correspoding symmetrical matrix) and
let $\Pi =\{ \alpha_{0}, \alpha_{1},\ldots, \alpha_{r}\}$ be a system
of simple roots  for $\hat{g}$.
We assume that the roots $\Pi_0  =
\{ \alpha_{1}, \alpha_{2},\ldots, \alpha_{r}\}$ generate the system
$\Delta_+ (g)$ of positive roots of the corresponding finite-dimensional
Lie algebra $g$. The quantum deformation $U'_{q}(\hat{g})$ is an
 associative algebra over the ring of formal power series ${\bf C}[[q-1]]$
  with generators $e_{\pm \alpha_{i}},\; k_{\alpha_{i}}^{\pm 1}=
q^{\pm h_{\alpha_{i}}},\; (i=1,2,\ldots,r)$, and the defining relations
\bn
[k_{\alpha_{i}}^{\pm 1}, k_{\alpha_{j}}^{\pm 1}]=0,\hsp
k_{\alpha_{i}}e_{\pm \alpha_{j}}= q^{\pm (\alpha_{i},\alpha_{j})}
e_{\pm\alpha_{j}} k_{\alpha_{i}},
\label{not.1}\ed
\bn
[e_{\alpha_{i}}, e_{-\alpha_{j}}] = \delta_{ij}\frac{k_{\alpha_{i}} -
k_{\alpha_{i}}^{-1}}{q-q^{-1}},
\label{not.2}\ed

\bn
(ad_{q'}e_{\pm\alpha_{i}})^{1-a_{ij}} e_{\pm\alpha_{j}}=0 \hsp
{\rm for}\;\,\,\, i\neq j, \:q'=q,q^{-1},
\label{not.3}
\ed
where $(ad_{q}e_{\alpha})e_{\beta}$ is a q-commutator:
\bn
(ad_{q}e_{\alpha})e_{\beta}\equiv
{[e_{\alpha},e_{\beta}]}_{q}=e_{\alpha}e_{\beta}-q^{(\alpha,\beta )}
e_{\beta}e_{\alpha}
\label{not.4}\ed
and  $(\alpha ,\beta )$ is a scalar  product of the roots $\alpha$
and $\beta$: $(\alpha_{i},\alpha_{j})=a_{ij}^{sym}$.

The condition $c=0$ is also imposed in ${U'}_{q}(\hat{g})$, where $c$ can be
 defined by the relation
$ q^c=k_{\a_0}^{n_0}k_{\a_1}^{n_1}\cdots k_{\a_r}^{n_r}$ with $n_0=\sum_{i\geq
 1}n_i$, $\sum_{i\geq 1}n_i\a_i=\theta$, $\theta$ being the highest root of
 $\Delta_+(g)$.

We define a comultiplication in ${U'}_{q}(\hat{g})$ by the formulas
\bn
\Delta (k_{\alpha_{i}})=k_{\alpha_{i}}\otimes k_{\alpha_{i}},\;
\label{Delta}
\label{not.5}\ed
\bn
\Delta (e_{\alpha_{i}}) = e_{\alpha_{i}}\otimes 1+
k^{-1}_{\alpha_{i}}\otimes e_{\alpha_{i}},
\;\:\Delta (e_{-\alpha_{i}}) = e_{-\alpha_{i}}\otimes
k_{\alpha_{i}} + 1 \otimes e_{-\alpha_{i}}
\label{Delta1}
\label{not.6}\ed

We  denote by a symbol $(^{*})$ an antiinvolution in $U_{q}(\hat{g})$,
defined as
$(k_{\alpha_{i}})^{*}=k_{\alpha_{i}}^{-1},\;$
$(e_{\pm\alpha_{i}})^{*}=e_{\mp\alpha_{i}},\;$
$(q)^{*}=q^{-1}$.

An algebra $U_{q}(\hat{g})$ is an associative algebra
over the ring of formal power series ${\bf C}[[q-1]]$
  with generators $e_{\pm \alpha_{i}},\; k_{\alpha_{i}}^{\pm 1}=
q^{\pm h_{\alpha_{i}}},\; (i=1,2,\ldots,r);\; d $, with the relations
(~\ref{not.1})--(~\ref{not.3}) and
\bn
[d,k_{\alpha_{i}}^{\pm 1}]=0,\hsp [d,e_{\pm \alpha_{i}}]= \pm \delta_{i,0}
e_{\pm \alpha_{i}}
\label{not.7}
\ed
Algebra $U_{q}(\hat{g})$ is also a Hopf algebra with the comultiplication
 (~\ref{not.5})--(~\ref{not.6}) and
$$ \Delta(d)=d\otimes 1+1\otimes d.$$

We use also the following standard notations:
$$exp_{q}(x) := 1 + x + \frac{x^{2}}{(2)_{q}!} + \ldots +
\frac{x^{n}}{(n)_{q}!} + \ldots = \sum_{n\geq 0} \frac{x^{n}}{(n)_{q}},$$

$$(a)_{q}:=\frac{q^{a}-1}{q-1},\,\,\,\,\,\,\,\,\,\,\,\,
[a]_{q}:=\frac{q^{a}-q^{-a}}{q-q^{-1}},\,\,\,\,\,\,\,\,\,\,\,\,
q_{\alpha}:=q^{-(\alpha,\alpha)}$$
and

$$k_\g=q^{h_\g}=k_{\a_0}^{n_0}\cdots k_{\a_r}^{n_r}$$
if $\g=n_0\a_0+\ldots n_r\a_r$.

\setcounter{equation}{0}
\section{Cartan-Weyl basis and the universal R-matrix for
quantum affine algebras}
Let $\Delta _{+}:=\D (\hat{g})$ be the system of all positive roots
of $\hat{g}$ with respect to $\Pi$.

It turns out that a
 procedure of the construction of the quantum Cartan-Weyl basis
 has to be in agreement with the choise of normal ordering in the
 root system \D. Let us recall the definition of normal order in
 \D \cite{AST}.
\begin{definition}
We say that the system \D is in normal (or convex) ordering if its roots
are totally ordered in the following way: ({\it i}) all multiple roots follow
each other in an arbitrary order; ({\it ii}) each nonsimple root
$\alpha +\beta \in\,$ \D, where $\alpha \neq \lambda\beta$ has to be
posed between $\alpha$ and $\beta$.
\end{definition}
    Fix some normal ordering in $\D =\Delta_+^{re}\cup\Delta_+^{im}$,
satisfying an additional condition:
     \bn
     \alpha_{i} +n\delta  <k\delta  <  (\delta  -\alpha_{j}  )
+l\delta \label{a}
     \ed
     for any simple roots $\alpha_{i},\alpha_{j}\in \D(g)$ , $
l,n \geq 0$, $k>0$ .  Here $\delta$ is a minimal
positive imaginary root.  Apply the following
    inductive procedure for the     construction
    of     real root      vectors $e_{\pm \gamma},$
 ,$\gamma  \in \Delta_+^{re}$ ,  starting from simple
root  vectors of  $\D$.

Let $\gamma \in \Delta_+^{re}$ be a real root and $\alpha ,\ldots ,
\gamma ,\ldots ,\beta $
be a minimal subset restricting  $\gamma\; (\gamma =\alpha +
\beta )$. Then we set
\bn
e_{\gamma}:={[e_{\alpha},e_{\beta}]}_{q},\;\;
e_{-\gamma}:={[e_{-\beta},e_{-\alpha}]}_{{q}^{-1}} \label{a44}
\ed
if $e_{\pm\alpha}$ and $e_{\pm\beta}$ have already been constructed.

 When  we   get   the
imaginary root  $\delta$  ,  we  stop for a moment and use the
following formulas:
\bn
{e_{\delta}^{(i)}}=\varepsilon_1(\a_i)
[e_{{\alpha}_{i}}, e_{\delta -{\alpha}_{i}}]_{q} \ ,
\label{C2}
\ed \
\bn
e_{\alpha_{i}+m\delta}=\varepsilon_n(\a_i)(-[(\alpha_i,\alpha_i)]_q)^{-m}
{(ad \; e_{\delta}^{(i)})}^{m}e_{{\alpha}_{i}} \ ,
\label{C3}
\ed \
\bn
e_{\delta -\alpha_{i}+m\delta}=\varepsilon_n(\a_i)
([(\alpha_i,\alpha_i)]_q)^{-m}
{(ad\; e_{\delta}^{(i)})}^{m}
e_{\delta -\alpha_{i}}  \ ,
\label{C4}
\ed \
\bn
{e'}_{m\delta}^{(i)}=\varepsilon_n(\a_i)
[e_{\alpha_{i}+(m-1)\delta},e_{\delta -\alpha_{i}}]_{q} \ ,
\label{C5}
\ed \\
(for $m > 0$), where $(ad\,x)y=[x,y]$ is a usual commutator,
$\varepsilon_n(\alpha_i)=(-1)^{n\theta(\alpha_i)}$,
 and $\theta :\Pi_0 =\{\alpha_1,\ldots  ,\alpha_r\} \rightarrow \{ 0,1\}$
is chosen in such a way that
     \bn
(\alpha_i,\alpha_j) \neq 0\;\;\;\;\longrightarrow\;\;\;\;
\theta(\alpha_i)\neq \theta(\alpha_j)
     \ed

  Then we use the inductive procedure again to obtain other
real root vectors $e_{\gamma +n\delta}$, $e_{\delta -\gamma +n
\delta}$ ,  $\gamma \in \D(g)$. We come to the end by defining
imaginary root     vectors     $e_{n\delta}^{(i)}$     through
intermediate vectors  ${e'}_{n\delta}^{(i)}$  by  means  of   the
following (Schur) relations:
\bn
(q-q^{-1})E^{(i)}(z)= \log \left( 1+(q-q^{-1}){E'}^{(i)}(z)\right)
\label{C6}
\ed
where $E^{(i)}(z)$ and ${E'}^{(i)}(z)$ are generating functions for
$e_{n\d}^{(i)}$ and for ${e'}^{(i)}_{n\d}$:
$$E^{(i)}(z)=\sum_{n \geq 1}e_{n\d}^{(i)}z^{-n},$$
$${E'}^{(i)}(z)=\sum_{n \geq 1}{e'}^{(i)}_{n\d}z^{-n}.$$
%%%%%%%%%%%%%%%%%%%%%%%%%%%%%%%%%%%%%%%%%%%%%%%%%%%%%%%%%%%%%
The root vectors of negative roots are obtained by the Cartan involution
 $(*)$: \
\bn
e_{-\gamma}=(e_{\gamma})^{*}
\label{C12}
\ed\
for $\gamma  \in \Delta_{+}(\hat{g})$.

An explicit expression for the universal $R$-matrix $\R$ (for the definition
see \cite{Dr1}) for quantum
nontwisted affine algebras was given in \cite{KT2}. Namely, for a fixed normal
ordering of $\Delta_{+}(\hat{g})$, satisfying (\ref{a}), we can present
$\R$ as follows:
\bn
\R =\overline{\R}_{re}^+\overline{\R}_{im}\overline{\R}_{re}^-\K
\ed
Here $ \K =q^{\sum_{i,j} d_{ij}h_i\otimes h_j+(c\otimes d+d\otimes c)}$,
     where
$d_{ij}$ is an inverse to  symmetrical Cartan
matrix $({b}_{ij}^{sym})$, $i,j= 1,\ldots r$ of underlying finite dimensional
 Lie algebra $g$,
\bn
   \overline{\R}_{re}^+= \prod_{\gamma\in{\Delta}_{+}^{re},\gamma <\delta}
     ^{\rightarrow}R_{\gamma},\hsp
   \overline{\R}_{re}^-= \prod_{\gamma\in{\Delta}_{+}^{re},\gamma >\delta}
     ^{\rightarrow}R_{\gamma} \label{R1}
\ed
where
\bn
    R_{\gamma}=
 \exp_{q_{\gamma }}\left( a(\gamma )
 e_{\gamma}\otimes e_{-\gamma}\right) ,
\ed
$a(\gamma )$ is ${\bf C}(q)$- coefficient in the relation
\bn
[e_\gamma, e_{-\gamma}]=\frac{k_\gamma-k_\gamma^{-1}}{a(\g)}
\ed
and the order in the product (\ref{R1}) coincides with
a chosen for the construction of the Cartan--Weyl basis normal
ordering of $\D$, satisfying (\ref{a}). Finally,
\bn
\overline{\R}_{im}= \exp\left( \sum_{n>0}^{}c_{i,j}^n e_{n\delta}^{(i)}
\otimes e_{-n\delta}^{(j)}\right)
\ed
where $c_{i,j}^n$ is $(q-q^{-1})$ times an inverse matrix to
\bn
\frac{[n(\a_i,\a_j)]_q}
{n}.
\ed

One can rewrite the general expression (\ref{R1}) for the
universal $R$-matrix in more symmetric form
 using the following property of an element $\K$
 in (\ref{R1}) (see \cite{KT1}):

$$ (e_\g \otimes 1)\K =\K (e_\g\otimes k_\g^{-1}),$$
 $$ (1\otimes e_{-\g})\K =\K (k_\g\otimes e_{-\g})$$

for any root vector $e_\g$, $\g \in \D$.

Let us define the following elements $\xi_{\pm\g}$, $\g\in \Delta_+^{re}$ and
 $x^{(i)}_{k}$, $i=1,\ldots ,r$, $k\in {\bf Z}$:

\bn
 \xi_{\g}=e_{\g},\;\;\;\xi_{-\g}=e_{-\g}  \hsp {\rm for}\;\; \g < \d,
\label{R2}
\ed
\bn
 \xi_{\g}= e_\g k_\g ,\;\;\; \xi_{-\g}=k_\g^{-1}e_{-\g}\hsp {\rm for}
\;\; \g >\d
\label{R3}
\ed
\bn
x^{(i)}_{\pm k}=e^{(i)}_{\pm k\d}\;\;\;{\rm for}\;\;k>0,\hsp
x^{(i)}_{0}=h_{\a_i}.
\label{R4}
\ed
Here $\d$ is again a minimal imaginary root.

In these notations the universal $R$-matrix for $U_q(\hat{g})$ can be written
 as

\bn
\R =\R^{+}_{re}\R_{im}\R^{-}_{re}
\label{R5}
\ed
where
\bn
\R^{+}_{re}= \prod_{\gamma\in{\Delta}_{+}^{re},\gamma <\delta}
     ^{\rightarrow}\exp_{q_\g}\left(a(\g)\xi_\g\otimes\xi_{-\g}\right),
\label{R6}
\ed
\bn
\R^{-}_{re}= \prod_{\gamma\in{\Delta}_{+}^{re},\gamma >\delta}
     ^{\rightarrow}\exp_{q_\g}\left(a(\g)\xi_\g\otimes\xi_{-\g}\right),
\label{R7}
\ed
\bn
\R_{im}= \exp\left(\sum_{n\geq 0}\sum_{i,j=1}^r c_{i,j}^n x_i\otimes x_{-j}
\right)\cdot q^{(c\otimes d +d\otimes c)}
\label{R8}
\ed
and
\bn
\left(c^n\right)_{i,j}=(q-q^{-1})\cdot {\left(\frac{[n(\a_k,\a_l)]_q}{n}
\right)^{-1}}_{i,j}
\label{R9}
\ed
for $n>0$ and
\bn
\left(c^0\right)_{i,j}=\log q\cdot{\left( (\a_k,\a_l)\right)^{-1}}_{i,j}
\label{R10}
\ed

The decomposition (\ref{R5}) produces the decomposition of the universal
 $R$-matrix $R$ for ${U'}_q(\hat{g})$:
\bn
R=R^+R^0R^-
\label{R11}
\ed
where $R^\pm =\R^\pm_{re}$ and $R^0=\R_{im}\cdot q^{-(c\otimes d+d\otimes c)}$.
%%%%%%%%%%%%%%%%%%%%%%%%%%%%%%%%%%%%%%%%%%%%%%%%%%%%%%%%%

In the following we use the notation of "finite vacua" representation of
 ${U'}_q(\hat{g})$ for a representation $V$ of ${U'}_q(\hat{g})$ whose
 restriction to $U_q(g)$ ($U_q(g)$ is generated be $e_{\pm\a_i},
k_{\a_i}^{\pm 1}$, $i=1,\ldots r$) is a highest weight module generated by
highest weight vector $v_0$.

We say that a vector $v\in V$ has a weight $\l (v)\in
h^*$ ($h$ is Cartan subalgebra of $g$) if
$k_{\a_i}v=\pm q^{(\l(v),\a)}v$ for $i=1,2,\ldots ,r$.
There is a natural partial ordering in the space $h^*$
of weights: $\l <\m$ if $(\m -\l ,\a_i)\in {\bf N}$,
$i=1,\ldots r$. This partial ordering induces a
partial ordering for weight vectors in finite vacua
representations of ${U'}_q(\hat{g})$: $v<u
\Leftrightarrow \l(u)<\l(v)$. We assume also that a
finite vacua representation $V$ is equiped with a
basis of weight vectors $v_i, i=0,1,\ldots $.

For a given finite vacua representation $V$ of ${U'}_q(\hat{g})$
 by means of the grading automorphism $D_z:{U'}_q(\hat{g})\rightarrow
 {U'}_q(\hat{g})$:
$$ D_z(e_{\pm\a_i})=z^{\pm \d_{i,0}}e_{\pm\a_i}, \hsp
D_z(k_{\a_i})=k_{\a_i}$$
one can define an evaluation representation (in homogenious gradation) $V(z)$
 of ${U'}_q(\hat{g})$: $\;\rho_{V(z)}=\rho_VD_z$.

The following theorem is a direct consequence of the presentation (\ref{R11})
 of universal $R$-matrix for ${U'}_q(\hat{g})$.
\begin{theorem}
For any two finite vacua representations $V$ and $W$
the decomposition (\ref{R11}) produces generalized
 Gauss decomposition of trigonometric $R$-matrix in
 $V(x)\otimes W(y)$:
\bn
\rho_{VW(z)}(R)=
\rho_{VW(z)}(R^+)
\rho_{VW(z)}(R^0)
\rho_{VW(z)}(R^-),
\label{R11a}
\ed
 $$\rho_{VW(z)}=\rho_{V(x)}\otimes\rho_{W(y)},\hsp z=x/y,$$
with
\bn
\rho_{VW(z)}(R^+)=1+\sum_{v_i\leq v_j,v_k\leq
v_l,\\(i,j)\neq (k,l)} a_{ij,kl}(z)e_{i,j}\otimes
e_{k,l}, \label{R12} \ed \bn
\rho_{VW(z)}(R^-)=1+\sum_{v_i\geq v_j,v_k\geq
 v_l,\\(i,j)\neq (k,l)} a_{ij,kl}(z)e_{i,j}\otimes
 e_{k,l}, \label{R13} \ed \bn
\rho_{VW(z)}(R^0)=\sum_{\l(v_i)=\l(v_j),\l(v_k)=\l(v_l)}
a_{ij,kl}(z)e_{i,j}\otimes e_{k,l}
\label{R14}
\ed
where the r.h.s. of (\ref{R12})--(\ref{R14}) are analitical functions of $z$
in the neibourhood of $0$.

The dependence of $R$ from $z$ is given by an action of $D_z\otimes 1$ on the
factors of $R$ in (\ref{R11}).
\label{Gauss}
\end{theorem}
\begin{corollary}
Let all the weight spaces of $V$ and $W$ are one
 dimensional (as it takes place for $sl_2$, for
  instance).

  Then the decomposition (\ref{R11a}) is exactly the
  Gauss decomposition of trigonometric $R$-matrix in
  $V(x)\otimes W(y)$.

   \label{corollary}
  \end{corollary}

\setcounter{equation}{0}
\section{Finite vacua representations of ${U'}_q(\widehat{sl}_2)$}
Let us first apply the procedure (\ref{a44})--(\ref{C12}) of the construction
 of the Cartan--Weyl basis for the algebra $U_q(\widehat{sl}_2)$.

Let $\alpha$ and $\beta=\delta -\alpha$ are simple roots for the affine
algebra $\widehat{sl_{2}}$ then $\delta=\alpha+\beta$ is a minimal
imaginary root.  We fix the following normal ordering in $\Delta_+$: \
\bn
\alpha,\;\alpha+\delta, \alpha+2\delta\ldots ,
\delta,\; 2\delta, \ldots , \ldots ,
\b +2\d ,\b +\d ,\b \ .
\label{A1}
\ed \
( another normal ordering is an inverse to (\ref{A1})).
 Following  (\ref{a44})--(\ref{C12}) we put
\bn
e'_{\delta}=e_{\delta}=
[e_{\alpha},e_{\b}]_{q} \ ,
\label{A2}
\ed

\bn
e_{\alpha+l\delta}=(-1)^{l}
\left( [(\alpha ,\alpha )]_{q}\right)^{-l}
(ad\ {e'}_{\delta})^{l}e_{\alpha} \ ,
\label{A3}
\ed

\bn
e_{\b+l\delta}=
\left( [(\alpha ,\alpha )]_{q}\right)^{-l}
(ad\ {e'}_{\delta})^{l}e_{\b} \ ,
\label{A4}
\ed

\bn
{e'}_{l\delta}=
[e_{\alpha+(l-1)\delta},e_{\b}]_{q}
\label{A5}
\ed  \\
and, finally,
\bn
(q-q^{-1})E(z)= \log \left( 1+(q-q^{-1}){E'}(z)\right)
\label{A6}
\ed
where $E(z)$ and ${E'}(z)$ are generating functions for
$e_{n\d}$ and for ${e'}_{n\d}$:
$$E(z)=\sum_{n \geq 1}e_{n\d}z^{-n},$$
$${E'}(z)=\sum_{n \geq 1}{e'}_{n\d}z^{-n}.$$
The negative root vectors are given by the rule
$e_{-\gamma}$ = $e_{\gamma}^{*}$.

It is well known \cite{J1} that there exists natural algebra epimorphism
 ${U'}_q(\widehat{sl}_2)$ $\rightarrow$ $U_q(sl_2)$:
\bn
e_\a\rightarrow e_\a,\hsp e_{\d -\a}\rightarrow e_{-\a}
\label{A6a}
\ed
and thus any highest weight representation $V$ of $U_q(sl_2)$ can be
extended to a finite vacua representation $V(x)$ of ${U'}_q(\widehat{sl}_2)$.
 Let $V_\l$ be Verma module for $U_q(sl_2)$ with a highest weight $\l$ and
highest weight vector $v_0$.
The action of generators $e_{\pm\a}$ and $k_\a$ of $U_q(sl_2)$ can be given
 in the basis $v_k, k\geq 0$ by the relations
\bn
e_\a (v_k)= [k]_q[\l -k]_qv_{k-1},\hsp e_{-\a}(v_k)= v_{k+1},
\label{A7}
\ed
\bn
k_\a (v_k)= q^{\l -2k}v_k,
\label{A8}
\ed
or, in matrix notations,
\bn
e_\a =\sum_{k\geq 0}[k]_q[\l -k]_qe_{k-1,k},\hsp
e_{-\a}=\sum_{k\geq 0}e_{k+1,k}
\label{A9}
\ed
\bn
k_\a =\sum_{k\geq 0}q^{\l -2k}e_{k,k}
\label{A10}.
\ed

Let $V_\l(-x)$ be corresponding finite vacua representation
 of ${U'}_q(\widehat{sl}_2)$ (we put the sign just for convenience of further
 notations). According to (\ref{A6a}), the structure of
 $V_\l(-x)$ is completely defined by
(\ref{A9})--(\ref{A10}) and the relation
$$e_\b=-xe_{\a}$$.
 One can easily check from (\ref{A2})
that the action of generators $e_{\pm\d}$ is given by the formula
\bn
e_{\pm\d}=x^{\pm 1}\sum_{k\geq 0}\left( q^{\mp 2}[k]_q[\l -k+1]_q
-[k+1]_q[\l -k]_q
\right)e_{k,k}
\label{A11}
\ed
and by (\ref{A3})--(\ref{A5}) we have
\bn
e_{\a +m\d}=x^m\sum_{k\geq 0}q^{-\l +2m-2}[k]_q[\l -k]_qe_{k-1,k,}
\label{A12}
\ed
\bn
e_{\b +m\d}=- x^{m+1}\sum_{k\geq 0}q^{-\l +2m}e_{k+1,k},
\label{A13}
\ed
and
$$
{e'}_{m\d}= x^m \sum_{k\geq 0}\left(
q^{(-\l +2k-2)(m-1)-2}[k]_q[\l -k+1]_q-\right.$$
\bn
\left.- q^{(-\l +2k)(m-1)}
[k+1]_q[\l -k]_q
 \right)e_{k,k}
\label{A14}
\ed
The relations (\ref{A12})--(\ref{A12}) may be rewritten in a more compact form:
\bn
e_{\a +m\d}=x^mk_\a^{-m}e_\a,
\label{A15}
\ed
\bn
e_{\b +m\d}=-e_{-\a}x^{m+1}k_\a^{-m},
\label{A16}
\ed
and also
\bn
e_{-(\a +m\d)}=e_{-\a}x^{-m}k_\a^m,
\label{A17}
\ed
\bn
e_{-(\b +m\d)}=-x^{-(m+1)}k_\a^me_\a
\label{A18}
\ed
Then we should write down the generating function
$E'(z)=\sum{e'}_{m\d}z^{-m}$:
$$E'(z)=\sum_{m\geq 1}\sum_{k\geq 0}x^mz^{-m}
\left( q^{(-\l +2k-2)(m-1)-2}[k]_q[\l -k+1]_q
-\right.$$
\bn
\left.
-q^{(-\l +2k)(m-1)}
[k+1]_q[\l -k]_q
\right)e_{k,k}
\label{A19}
\ed
and take the logarithm of $1+(q-q^{-1})E'(z)$. Formal elementary manipulations
 with logarithms give  the following answer:
$$
(q-q^{-1})E(z)=
\log (1-q^{-2}k_\a^{-1}
x/z)+\log (1-k_\a^{-1} x/z)-$$
\bn
-\log (1-q^\l x/z)-\log (1-q^{-\l-2}x/z)
\label{A20}
\ed
and thus the action of the imaginary root vectors is given by the following
 formulas:
\bn
e_{m\d}= \frac{q^{\l m}+q^{(-\l -2)m}-k_\a^{-m}-q^{-2m}k_\a^{-m}}
{m(q-q^{-1})}x^m
\label{A21}
\ed
\bn
e_{-m\d}= \frac{-q^{-\l m}-q^{(\l +2)m}+k_\a^{m}+q^{2m}k_\a^{m}}
{m(q-q^{-1})}x^{-m}
\label{A22}
\ed
\setcounter{equation}{0}
\section{$R$-matrix in tensor product of $U_q(sl_2)$-Verma modules}
The universal $R$-matrix for $U_q(\widehat{sl}_2)$ has the following form
 \cite{KT2}:
\[
{\cal R}=\left(\prod_{n\geq 0}^{\rightarrow} \exp_{q_{\alpha}}
\left( (q-q^{-1})e_{\alpha+n\delta} \otimes e_{-\alpha-n\delta}
\right)\right)\cdot
\]
\[
\exp\left( \sum_{n>0}(q-q^{-1})\frac{n(e_{n\delta}
\otimes e_{-n\delta})}{[n(\a ,\a)]_q}\right)\cdot
\]
\bn
\left(\prod_{n\geq 0}^{\leftarrow}\exp_{q_{\alpha}}\left( (q-q^{-1})
e_{\beta+n\delta}\otimes e_{-\beta-n\delta}\right)\right)\cdot {\cal K}
,
\label{B1}
\ed \\
where the order on n is direct in the first product and it is inverse
in the second one. Factor ${\cal K}$ is defined by the  formula:
\bn
{\cal K}=q^{{\frac{h_\a\otimes h_\a}{(\a ,\a)}}}\cdot
q^{(c\otimes d+d\otimes c)}
\label{B2}
\ed
Applying the transformations (\ref{R2})--(\ref{R4}) we obtain from (\ref{B1})
the following expression for the universal $R$-matrix for
 ${U'}_q(\widehat{sl}_2)$:
\[
 R=\left(\prod_{n\geq 0}^{\rightarrow} \exp_{q_{\alpha}}
\left( (q-q^{-1})e_{\alpha+n\delta} \otimes e_{-\alpha-n\delta}
\right)\right)\cdot
\]
\[
q^{{\frac{h_\a\otimes h_\a}{(\a ,\a)}}}\cdot
\exp\left( (q-q^{-1})\sum_{n>0}\frac{n(e_{n\delta}
\otimes e_{-n\delta})}{[n(\a ,\a)]_q}\right)\cdot
\]
\bn
\left(\prod_{n\geq 0}^{\leftarrow}\exp_{q_{\alpha}}\left( (q-q^{-1})
e_{\beta+n\delta}k_\a^{-1}\otimes k_\a e_{-\beta-n\delta}\right)\right) .
\label{B3}
\ed \\
Let now $V_\l$ and $W_\m$ be $U_q(sl_2)$-Verma modules
(or their factormodules) with highest weights $\l$ and
$\m$.  Substituting expression
(\ref{A15})--(\ref{A18}) and (\ref{A21})--(\ref{A22})
into (\ref{B3}) we obtain the following expression for
 the $R$-matrix $R_{VW}(z)$ in $V_\l (x)\otimes W_\m
 (y)$ ($z=x/y$):
$$R_{VW}(z)=R^+_{VW}(z)R^0_{VW}(z)R^-_{VW}(z)$$
where
 \bn
R^+_{VW}(z) =\left(\prod_{n\geq 0}^{\rightarrow} \exp_{q^{-2}}
\left( (q-q^{-1})z^nk_\a^{-n}e_{\alpha} \otimes e_{-\alpha}k_\a^n
\right)\right) ,
\label{B4}
\ed
\bn
R^-_{VW}(z)=
\left(\prod_{n\geq 1}^{\leftarrow}\exp_{q^{-2}}\left( (q-q^{-1})z^n
e_{-\a}k_\a^{-n}\otimes k_\a^n e_{\a}\right)\right) ,
\label{B5}
\ed
\bn
R^0_{VW}(z)=
q^{{\frac{h_\a\otimes h_\a}{2}}}\cdot
\label{B6}
\ed
%%%%%%%%%%%%%%%%%%%%%%%%%%%%%%%%%%%%%%
\[
\exp\left( \sum_{n>0}\frac{z^n}{n}\frac{\left(
q^{\l n}+q^{(-\l -2)n}-k_\a^{-n}-q^{-2n}k_\a^{-n}\right)\otimes
 \left( -q^{-\mu n}-q^{(\mu
+2)n}+k_\a^{n}+q^{2n}k_\a^{n}\right)} {q^{2n}-q^{-2n}}
\right)
\]
Let $u$ be the following diagonal matrix in $V_\l\otimes W_\m$:
\bn
u=z\cdot k_\a\otimes k_\a^{-1}
\label{B7}
\ed
and
\bn
v_+=(q-q^{-1})e_\a\otimes e_{-\a},\hsp
v_-=(q-q^{-1})e_{-\a}\otimes e_{\a}.
\label{B8}
\ed
Then the expansion of rhs of (\ref{B4}) over the powers of
$v_+$ looks as follows ($p=q^{-2}$):
\[
R^+_{VW}(z) =
1+\sum_{k\geq 0}(up)^kv_++
\frac{1}{(2)_p}\sum_{k\geq 0}(k+1)_p(up)^k
v_+^2+
\]
\[
+\ldots\frac{1}{(n)_p!}\sum_{k\geq
0}\frac{(k+1)_p\cdots (k+n-1)_p}{(n-1)_p!} (up)^k
v_+^k
+\ldots
\]
or, in other terms,
\[
R^+_{VW}(z) =1+\frac{1}{1-up}v_++
\frac{1}{(2)_p}\cdot\frac{1}{(1-up)(1-up^2)}
v_+^2+
\]
\bn
\ldots +
\frac{1}{(n)_p!}\cdot\frac{1}{(1-up)\cdots (1-up^n)}v_+^n+
\ldots
\label{B9}
\ed
One can interprete an expansion (\ref{B9}) as (ordered) $p$-exponent:
\bn
R^+_{VW}(z) =:\exp_p \frac{q-q^{-1}}{1-up}\cdot\left(
e_\a\otimes e_{-\a} \right) :
\label{B10}
\ed
if
$:\exp_ta(p)\cdot b(p):$ means
\[ :\exp_ta(p)b(p):=
1+ a(p)b(p)+\frac{1}{(2)_t!}a(p)a(p^2)\cdot
b(p)b(p^2)+ \]
\bn \ldots +\frac{1}{(n)_t!}a(p)\cdots
a(p^n)\cdot b(p)\cdots b(p^n)+\ldots
\label{B11}
\ed
Analogously,
\[
R^-_{VW}(z) =1+\frac{up}{1-up}v_-+
\frac{p}{(2)_p}\cdot\frac{up}{(1-up)}\frac{up^2}{(1-up^2)}
v_-^2+
\]
\bn
\ldots +
\frac{p^{\frac{n(n-1)}{2}}}{(n)_p!}\cdot\frac{up}{(1-up)}\cdots
\frac{up^n}{(1-up^n)}v_-^n+
\ldots
\label{B12}
\ed
or
\bn
R^-_{VW}(z) =:\exp_{p^{-1}}
\frac{(q-q^{-1})up}{1-up}\cdot \left( e_{-\a}\otimes
e_{\a} \right) :
\label{B13}
\ed
The expression
(\ref{B6}) for $R^0_{VW}(z)$ we can also transform in
the following way:
\bn R^0_{VW}(z)=f_{\l ,\m}(z )\cdot
q^{\frac{h_\a\otimes h_\a}{2}}\cdot
 R^{(')}_{VW}(z)\cdot R^{('')}_{VW}(z)
\label{B14}
\ed
 where
\bn f_{\l ,\m}(\l )=\exp  \sum_{n\geq 1}
\left( (q-q^{-1})
\frac{[\l n]_q[\m n]_q}{[2n]_q}
\right)\frac{z^n}{n} ,
\label{B15}
\ed
\bn
R^{(')}_{VW}(z)=\exp  \sum_{n\geq 1}
\left( (q^{-\l n}-k_\a^{-n})\otimes
q^{-n}\frac{[\m n]_q}{[n]_q}+
q^n\frac{[\l n]_q}{[n]_q}
\otimes (k_\a^n-q^{\m n})\right)
\frac{z^n}{n}
\label{B16}
\ed
and
\bn
R^{('')}_{VW}(z)=\exp  \sum_{n\geq 1}\left(
\frac{q^n+q^{-n}}{(q^n-q^{-n})}\cdot \left( q^{-\l n}- k_\a^{-n}\right)
\otimes \left( k_\a^n-q^{\m n}\right)\right)
\frac{z^n}{n}
\label{B17}
\ed

\setcounter{equation}{0}
\section{Discussions}
 Let us observe the formulas for $R$-matrix in tensor product of finite vacua
 representations $V_\l (x)\otimes W_\m (y)$
 of ${U'}_q(\widehat{sl}_2)$.
We present $R$ in a form of Gauss decomposition
$$R=R^+_{VW}(z)\cdot R^0_{VW}(z)\cdot R^-_{VW}(z)$$
with $R^\pm_{VW}(z)$  given by expressions (\ref{B9}) and (\ref{B12}) and
 $R^0_{VW}(z)$ given by (\ref{B14})--(\ref{B17}). One can easily see that only
 finite number of summands of rhs seria (\ref{B9}) and (\ref{B12})
act on a fixed weight vector of
 $V_\l \otimes W_\m $ and the matrix coefficients of $R^\pm_{VW}(z)$ are
 rational functions of $z=x/y$ with coefficients
depending on weights.  The number of poles of these
coefficients is unbounded for Verma modules and
 bounded for finite dimensional representations (the
 formulas are valid for all finite vacua
 representations).

The structure of $R^0_{VW}(z)$ is more trancendental. We express
 $R^0_{VW}(z)$ as a product
$$R^0_{VW}(z)=f_{\l ,\m}(z )\cdot
 q^{\frac{h_\a\otimes h_\a}{2}}
 \cdot
 R^{(')}_{VW}(z)\cdot R^{('')}_{VW}(z)$$
with the factors given by (\ref{B15})--(\ref{B17}).
The scalar factor $f_{\l ,\m}(z)$ comes from
 nonlinear equation
\bn
(\Delta\otimes Id){\cal R}={\cal
R}^{13}{\cal R}^{23}
\label{5.1}
\ed
 which distinguish $R$-matrices
originating from the universal $R$-matrix from other
solutions of the Yang-Baxter equations.
An equation (\ref{5.1}) is an origin of the following
 functional equation on $f_{\l ,\m}(z)$ \cite{Dr2}:
$$\frac{f_{\l ,\m}(q^2z)}{f_{\l ,\m}(q^{-2}z)}=
\frac{(1-q^{(\l -\m)}z)(1-q^{(\m -\l)}z)}
{(1-q^{(\l +\m)}z)(1-q^{-(\m +\l)}z)}$$
 and express $f_{\l ,\m}(z)$
 in terms of $q-gamma$ functions (for rational
variant see \cite{S}) or, more conveniently, in terms
of infinite products $(z;q)_\infty$ \cite{GR}, where
$$(z;q)_\infty =\prod_{k=0}^{\infty}(1-zq^k):$$
 \bn
f_{\l ,\m}(z)=\frac{(zq^{\l -\m-2};q^{-4})
(zq^{\m -\l-2};q^{-4})}{(zq^{\l +\m-2};q^{-4})
(zq^{-\l -\m-2};q^{-4})}
\label{5.2}
\ed
Presentation (\ref{5.2}) has sense for any $q\neq 1$
since the sum of weights in nominator and denominator
of r.h.s. of (\ref{5.2}) coincide.

Such a function was used in \cite{FR} as a simplest
solution of $q$-difference Knizhnik-Zamolodchikov
equation. The diagonal matrices $R^{(')}_{VW}(z)$ and
$R^{('')}_{VW}(z)$ have  rational entries for
 integer weights $\l$ and $\m$.  For instance, one can
 easily see that for $R^{(')}_{VW}(z)$ if notices that
 the coefficients $\frac{[\l n]_q}{[n]_q}$ are finite
polinomials (geometric progressions) in this case.
 But for noninteger weights $\l$ and $\m$ the entries
of $R^{(')}_{VW}(z)$ and  $R^{('')}_{VW}(z)$ are also
 can be expressed only in terms of $q-gamma$
functions analogously to (\ref{5.2}).
 This means that for generic Verma modules
$V_\l$ and $W_\m$ the trigonometric
$R$ matrix in $V_\l (x)\otimes W_\m (y)$ cannot be expressed  as a matrix
with entries of rational functions. However, the formulas
(\ref{B15})--(\ref{B17}) give possibility to express easily all the matrix
  coefficients in terms of basic  functions
$(z;q)_\infty$. Note that there is an analogous
presentation for rational $R$-matrices. It is
discussed in the forthcoming paper.

%{\bf References}


\begin{thebibliography}{99}
%\medskip
%\begin{itemize}
%%\vspace{0.5in}
%%\centerline{{\bf References}}
%%\vspace{0.5cm}

\bibitem[AST]{AST}
 Asherova, R.M., Smirnov, Yu.F., and Tolstoy, V.N. A description  of
some class of projection operators  for  semisimple  complex
Lie algebras. {\it Matem. Zametki 26} (1979), 15-25.

\bibitem[Ch1]{Ch1}
Cherednik, I.V., A new interpretation of Gelfand-Zetlin bases.
{\it Duke Math. J. 54} (1987), 563-577.

\bibitem[Ch2]{Ch2}
Cherednik, I.V., Quantum groups as hidden symmetries
of classical representation theory. {\it in
''Differential geometric methods of theoretical
Physics'' (A.I.Solomon Ed.),World
Scientific, Singapore} (1989),  47-54.

\bibitem[Dr1]{Dr}
 Drinfeld, V.G. A new realization of Yangians and quantized
 affine algebras. {\it Soviet Math. Dokl. 32} (1988), 212-216.

\bibitem[Dr2]{Dr1}
 Drinfeld, V.G. Quantum groups. {\it Proc. ICM-86 (Berkely USA) vol.1},
798-820. Amer. Math. Soc. (1987).

\bibitem[Dr3]{Dr2}
 Drinfeld, V.G. Hopf algebras and the quantum Yang--Baxter equation.
. {\it Soviet Math. Dokl. 32 }, (1985), 254-258.

\bibitem[FR]{FR}
 Frenkel, I.B., and Reshetikhin, N.Yu.
 Quantum Affine  Algebras and Holonomic Difference equations.
 {\it Commun. Math. Phys., 146} (1992), 1-60.

\bibitem[GR]{GR}
 Gasper, G., and Rahman, M. Basic Hypergeometric
Series. {\it Cambridge Univ. Press} (1990).

\bibitem[J1]{J1}
Jimbo, M. Quantum $R$-matrix for the generalized Toda system.
{\it Comm. Math. Phys. 102} (1986), 537-547.


\bibitem[J2]{J2}
 Jimbo, M. A $q$-difference analogue of $U(g)$ and the Yang-Baxter
equation. {\it Lett. Math. Phys. 10} (1985), 63-69.


\bibitem[KS]{KS}
Kulish, P.P. and Sklyanin, E.K. Quantum spectral transform method:
                  recent developments. {\it in ''Integrable quantum
 field theories'', LN in Physics 151} (1982), 61-119.

\bibitem[KRS]{KRS}
Kulish, P.P., Reshetikhin, N.Yu. and Sklyanin, E.K.
Yang--Baxter Equation and Representation Theory I.
{\it Let.Math.Phys. 5} (1981), 393-403.

\bibitem[KT1]{KT1}
Khoroshkin, S.M., and Tolstoy, V.N. Universal $R$-matrix
for quantized (super)algebras. {\it Commun. Math. Phys. 141} (1991), 599-617.

\bibitem[KT2]{KT2}
 Khoroshkin, S.M., and Tolstoy, V.N. The Universal $R$-matrix
 for quantum nontwisted
affine Lie algebras. {\it Funkz. Analiz i ego pril. 26:1} (1992), 85-88.

\bibitem[KT3]{KT3}
 Khoroshkin, S.M., and Tolstoy, V.N. The
Uniqueness Theorem for the Universal $R$-matrix.
{\it Lett. Math. Phys., 24} (1992), 231-244.

\bibitem[S]{S}
 Smirnov, F.A. Dinamical symmetries of massive integrable models
{\it J. of Modern Phys. A, 7 suppl. 1B} (1992), 813-838.

\bibitem[ZG]{yzz}
  Zhang, Y-Z., and Gould, M.D.,
On Universal $R$-matrix for quantized nontwisted rank 3 affine Lie algebras.
 {\it Lett. Math. Phys. 29} (1993), 19.

\end{thebibliography}
\end{document}